\documentclass[runningheads,a4paper]{llncs}

\usepackage{graphics}

\usepackage[colorinlistoftodos]{todonotes}

\usepackage{amsmath}
\usepackage{amsfonts}
\usepackage{amssymb}

\newtheorem{ex}{Example}

\usepackage{color}
\usepackage[colorinlistoftodos]{todonotes}
\usepackage{verbatim}
\usepackage{url}

\usepackage{tikz}
\usetikzlibrary{positioning,decorations.pathmorphing, decorations.pathreplacing,patterns,shapes,arrows}
\tikzstyle{vertex}=[circle,draw=black, fill=black!10,minimum size=20pt,inner sep=0pt, text centered]
\tikzstyle{dot}=[circle,draw=black, fill=black,minimum size=2pt,inner sep=0pt, text centered]
\tikzstyle{edge} = [draw,thick,-]
\tikzstyle{zig} = [draw,decoration = {zigzag,segment length = 3mm, amplitude = 1mm},decorate]
\newcommand{\defproblem}[3]{
  \vspace{2mm}
\noindent\fbox{
  \begin{minipage}{0.96\textwidth}
  #1\\
  {\bf{Input:}} #2  \\
  {\bf{Output:}} #3
  \end{minipage}
  }
  \vspace{2mm}
}

\def\dd{\mathinner{\ldotp\ldotp}} % dot dot

\newcommand{\match}{\simeq} %matching symbol
\newcommand{\els}{\hat} %EDS symbol
\newcommand{\el}{\xi} %EDF symbol
\newcommand{\sig}{\operatornamewithlimits{\sum}\limits}

\newcommand{\ticked}{\mathcal{TT}} %Sequence of EDF
\newcommand{\sz}{\|} %size

\newcommand{\ord}{\mathcal{O}} %Order symbol
\newcommand{\mystep}[1]{\noindent \textbf{#1}}
\newcommand{\mymat}{\mathbf}
\definecolor{mygray}{gray}{0.8} 
\usepackage[algo2e, noend, noline, algoruled]{algorithm2e}
  \renewcommand{\CommentSty}[1]{\textnormal{#1}\unskip}%%\SetDataSty{texttt}
  \SetKwComment{tcm}{\{}{\}}
  \SetCommentSty{mycommfont}
  \SetKw{KwAnd}{and}%
  \SetKw{KwOr}{or}%
  \SetKw{KwNot}{not}%
    \SetKw{KwIn}{in}%

\begin{document}
\mainmatter 
\title{Efficient Pattern Matching in Elastic-Degenerate Strings\footnote{This work was partially supported by the British Council funded INSPIRE Project}}
\titlerunning{Pattern Matching in Elastic-Degenerate Strings} %optional, in case that the title is too long; the running title should fit into the top page column

%% Please provide for each author the \author and \affil macro, even when authors have the same affiliation, i.e. for each author there needs to be the  \author and \affil macros

\author{Costas~S.~Iliopoulos
\and Ritu~Kundu
\and Solon~P.~Pissis
}
\institute{Department of Informatics,
    King's College London, London WC2R 2LS, UK \\
  \texttt{\{costas.iliopoulos, ritu.kundu, solon.pissis\}@kcl.ac.uk}
}
\authorrunning{\,C. Iliopoulos, \,R. Kundu, and \,S. Pissis} %mandatory. First: Use 
\date{}

\toctitle{Efficient Pattern Matching in Elastic-Degenerate Strings}
\tocauthor{Iliopoulos et. al}
\maketitle 
\begin{abstract}

In this paper, we extend the notion of gapped strings to \textit{elastic-degenerate strings}. An elastic-degenerate string can been seen as an ordered collection of $k>1$ seeds (substrings/subpatterns) interleaved by \textit{elastic-degenerate symbols} such that each elastic-degenerate symbol corresponds to a set of two or more variable length strings. 

Here, we present an algorithm for solving the pattern matching problem with (solid) pattern and elastic-degenerate text, running in $\ord(N+\alpha\gamma nm)$ time; where $m$ is the length of the given  pattern; $n$ and $N$ are the length and total size of the given elastic-degenerate text, respectively; $\alpha$ and $\gamma$ are small constants, respectively representing the maximum number of strings in any elastic-degenerate symbol of the text and the largest number of elastic-degenerate symbols spanned by any occurrence of the pattern in the text. The space used by the algorithm is linear in the size of the input for a constant number of elastic-degenerate symbols in the text; $\alpha$ and $\gamma$ are so small in real applications that the algorithm is expected to work very efficiently in practice.

\end{abstract}

%---------%---------%---------%---------%---------%---------%---------%--------%
%---------%---------%---------%---------%---------%---------%---------%--------%
%---------%---------%---------%---------%---------%---------%---------%--------%

\section{Introduction}
\label{sec:intro}

Uncertainty in sequential data (strings) can be characterised using various representations. One such representation is \textit{degenerate string} which is defined by the existence of one or more positions that are represented by sets of symbols from an alphabet $\Sigma$, unlike an accurate or certain (standard) string characterised by a single symbol at each position. For instance, \mbox{\scriptsize
$\bigl[\begin{smallmatrix}
\texttt{a} \\\texttt{b}
\end{smallmatrix} \bigr]
\texttt{a} \texttt{c}
\bigl[\begin{smallmatrix}
\texttt{b} \\\texttt{c}
\end{smallmatrix} \bigr]
\texttt{a}
\bigl[\begin{smallmatrix}
\texttt{a} \\ \texttt{b}  \\ \texttt{c}
\end{smallmatrix} \bigr]
$}is a degenerate string of length $6$ over $\Sigma = \{\texttt{a,b,c}\}$. 

A \textit{compound pattern}(or \textit{gapped pattern}) is another way to capture uncertainty -- it is a list of standard (simple) sub-patterns (or seeds) separated by variable length gaps defined by a list of intervals \cite{5ab9058d631145c29eb4abcd7346305a}. Simply, a compound pattern $P$ can be represented as follows \cite{Rahman2006}: $P = P_1 *^{a_1, b_1} P_2 *^{a_2, b_2} P_3 \cdots *^{a_{l-1}, b_{l-1}} P_l$
%\begin{center}$P = P_1 *^{a_1, b_1} P_2 *^{a_2, b_2} P_3 \cdots *^{a_{l-1}, b_{l-1}} P_l$\end{center}
where,  $*$ is a \textit{wildcard character} (also called \textit{don't care symbol} or \textit{hole}) that matches any character in a finite alphabet $\Sigma$; $\forall i \in [1..l]$ each sub-pattern $P_i $ is a string over $\Sigma$; and $\forall i \in [1..l-1]$ each pair $(a_i, b_i)$ represents the gap (minimum and maximum wildcard characters, respectively) between two consecutive subpatterns $P_i$ and $P_{i+1}$.

Here, we introduce another representation to encapsulate uncertainty in sequential data -- which we call \textit{elastic-degenerate strings} -- by extending and combining the idea of gapped patterns/strings and degenerate strings. An \textit{elastic-degenerate string} is a string such that at one or more positions, an \textit{elastic-degenerate symbol} can occur which is defined as a set of variable length substrings. Another way to visualise an elastic-degenerate string is to see it as an ordered collection of $k>1$ seeds (substrings) interleaved by elastic-degenerate symbols such that each elastic-degenerate symbol corresponds to a set of two or more variable length substrings. 
\mbox{\scriptsize
$\texttt{bc}
\begin{bmatrix}
\texttt{ab} \\ \texttt{aab} \\ \texttt{aca}
\end{bmatrix}
\texttt{ca}
\begin{bmatrix}
\texttt{abcab} \\ \texttt{cba}
\end{bmatrix}
\texttt{bb}$
} is an example of an elastic-degenerate string over $\Sigma = \{\texttt{a,b,c}\}$.

This generalisation of concept of \textit{degeneracy} is motivated by several important data mining problems which can be reduced to the core task of discovering occurrences of one or more patterns in a text that can best be described as an ordered collection of strings interleaved by sets of variable length strings. 

More specifically, in genomics an important class of problems is to study within-species genetic variation; the state of the art solutions for this class comprises of matching(\textit{mapping}) substrings (called \textit{reads}) to a longer genomic sequence(canonical \textit{reference genome} obtained through assembly). Owing to the high diversity among biologically relevant genomic regions in many organisms, the population level complexities can not be captured by the `linear' structure of a reference genome (see \cite{Liu2014}). Consequently, recent research trend has shifted towards using alternative representations of genomic sequence for population-based genome assembly (\cite{Huang01072013,Church2015,Dilthey2015,Maciuca2016}). One such representation that encodes a set of related genomes with variations in the reference genome itself (called Population Reference Genome in \cite{Maciuca2016}), can be seen as an elastic-degenerate string.

The problem of matching in the context of gapped strings has been studied extensively using combinatorial approaches (see \cite{Pissis2014} and references therein). %approaches~\cite{pmid11108467,Carvalho:2004:PAE:967900.967932,Carvalho2004,pmid16870937,pis:car:mar:sag:06,doi:10.1142/S0129054105003716,AlatabbiAHIR15,Zhang2006,Pissis2014,CrochemoreEtAl_2000_FindMotiWithGaps,ILN02MotifsPolyphonic}. 
However, a gapped string (which specifies the constraint on only the length of the gap between two consecutive seeds) differs from an elastic-degenerate string because the later precisely defines the possible  substrings (of varying lengths) that can exist between those consecutive seeds. This precise identification of allowed substrings in a gap makes the matching problem in the context of elastic-degenerate strings, algorithmically more challenging and computationally difficult.

In this paper, we not only formalize the concept of elastic-degenerate strings but also present an efficient - in terms of both, space and time - algorithm to solve the pattern matching problem in a given elastic-degenerate text. To the best of our knowledge, no other work, heretofore explores the problem accounting for \textit{elastic-degeneracy} in the text.
%We will generally be following \cite{AlatabbiAHIR15} for notations and definitions.

%The rest of the paper is organised as follows:
%In the next section, we introduce the basic definitions and establish the notions of elastic-degeneracy; the algorithmic tools required to build the solution are described in Section \ref{sec:tools}; next, we formally define the problem along with presenting the algorithm; the algorithm is analysed in Section \ref{sec:analysis}; Finally, the paper is concluded.
In the next section, we introduce the basic definitions and establish the notions of elastic-degeneracy that will be used in this paper. The algorithmic tools required to build the solution are described in Section \ref{sec:tools}. In Section \ref{sec:algo}, we formally define the problem along with presenting the algorithm. The algorithm is analysed in Section \ref{sec:analysis}.Finally, the paper is concluded in Section \ref{sec:conclusion}.

%---------%---------%---------%---------%---------%---------%---------%--------%
%---------%---------%---------%---------%---------%---------%---------%--------%
%---------%---------%---------%---------%---------%---------%---------%--------%
%%%%%%%%%%%%%%%% Section 2 %%%%%%%%%%%%%%%%%%%%%%%%%%%%%5
%---------%---------%---------%---------%---------%---------%---------%--------%

\section{Terminology and Technical Background} 
%
%\subsection{Definitions and Notation}

We begin with basic definitions and notations. We think of a \textit{string} $X$ of \textit{length} $n$ as an array $X[1\dd n]$,
where every $X[i]$, $1 \le i \le n$, is a \textit{letter} drawn from some fixed \textit{alphabet} $\Sigma$ of size $|\Sigma| = \mathcal{O}(1)$.
The \textit{empty string} is denoted by $\varepsilon$.
$\Sigma^*$ denotes the set of all strings over an alphabet $\Sigma$ including empty string $\varepsilon$.
A string $Y$ is a \textit{factor} of a string $X$ if there exist two strings $U$ and $V$, such that $X=UYV$. Hence, we say that there is an \textit{occurrence} of $Y$ in $X$, or simply, that $Y$ \textit{occurs in} $X$. The starting position of an occurrence, say $i$, is called \textit{head} of the occurrence and its ending position ($i + \vert Y \vert$ - 1) is called its \textit{tail}. Note that an empty string occurs at each position in a given string.

Consider the strings $X,~Y,~U$, and $V$, such that $X=UYV$.
If $U=\varepsilon$, then $Y$ is a \textit{prefix} of $X$.
If $V=\varepsilon$, then $Y$ is a \textit{suffix} of $X$.  

A \textit{degenerate symbol} $\tilde{\sigma}$ over an alphabet $\Sigma$ is a non-empty subset of $\Sigma$, i.e., $\tilde{\sigma} \subseteq \Sigma$ and $\tilde{\sigma} \neq \emptyset$. $\vert \tilde{\sigma} \vert$ denotes the size of the set and we have $1 \leq \vert \tilde{\sigma} \vert \leq  \vert \Sigma \vert$. A \textit{degenerate string} is built over the potential $2^{\vert \Sigma \vert} - 1$ non-empty sets of letters belonging to $\Sigma$.  In other words,  a degenerate string  $\tilde{X} = \tilde{X}[1 .. n]$, is a string such that every $\tilde{X}[i]$ is a degenerate symbol, $1\leq i\leq n$. 
If $\vert \tilde{x}[i] \vert = 1$, that is, $\tilde{X}[i]$ represents a single symbol of $\Sigma$, we say that $\tilde{X}[i]$ is a \textit{solid} symbol and $i$ is a \textit{solid position}. Otherwise $\tilde{X}[i]$  and $i$ are said to be a \textit{non-solid symbol} and a \textit{non-solid position}, respectively.
For example, \mbox{\scriptsize
$\bigl[\begin{smallmatrix}
\texttt{a} \\\texttt{b}
\end{smallmatrix} \bigr]
\texttt{a} \texttt{c}
\bigl[\begin{smallmatrix}
\texttt{b} \\\texttt{c}
\end{smallmatrix} \bigr]
\texttt{a}
\bigl[\begin{smallmatrix}
\texttt{a} \\ \texttt{b}  \\ \texttt{c}
\end{smallmatrix} \bigr]
$} is a degenerate string of length $6$ over $\Sigma = \{\texttt{a, b, c}\}$. 
A string containing only solid symbols will be called a \textit{solid string}. 
A \textit{conserved degenerate string} is a degenerate string where its number of non-solid symbols is upper-bounded by a fixed positive constant $k$. 

Now we give the terminology to build the concept of elastic-degeneracy by presenting the following definitions and examples.

\begin{definition}[Seed: $S$] A \textit{seed} $S$ is a (possibly empty) string over $\Sigma$ (i.e $S\in \Sigma^*$).
\end{definition}

\begin{definition}[Elastic-Degenerate Symbol: $\el$] An \textit{elastic-degenerate symbol} $\el$, over a given alphabet $\Sigma$, is a non-empty set of strings over $\Sigma$ (i.e $\el \subset \Sigma^*$ and $\el \neq \emptyset$). An elastic-degenerate symbol $\el$ is denoted by \mbox{\scriptsize
$\begin{bmatrix}
E_{1} \\ E_{2} \\ \vdots \\ E_{\vert \el \vert}
\end{bmatrix}$}, where each $E_i, ~1 \leq i \leq \vert \el \vert$ is a solid string. The \textit{minimum (maximum) length} of $\el$, represented as $\vert \el \vert_{min}$ ($\vert \el \vert_{max}$), is the length of the shortest (or longest) string in the set. 
\end{definition}

\begin{definition}[Elastic-Degenerate String: $\els{X}$] An \textit{elastic-degenerate string} $\els{X}$, over a given alphabet $\Sigma$, is a sequence $S_1 \el_1 S_2 \el_2 S_3 \dd S_{k-1}\el_{k-1} S_k$, where $S_i,~1 \leq i \leq k$ is a seed and $\el_i,~ 1 \leq i \leq k-1$ is an elastic-degenerate symbol. 
\end{definition}

An elastic degenerate string $\els{X}$ can be visualised as follows:
\begin{center}
$\els{X} = $\mbox{\scriptsize $S_1 
\begin{bmatrix}
E_{1,1} \\ E_{1,2} \\ \vdots \\ E_{1,\vert \el_1 \vert}
\end{bmatrix}
S_2
\begin{bmatrix}
E_{2,1} \\ E_{2,2} \\ \vdots \\ E_{2,\vert \el_2 \vert}
\end{bmatrix}
S_3\dd S_{k-1}
\begin{bmatrix}
E_{k-1,1} \\ E_{k-1,2} \\ \vdots \\ E_{k-1,\vert \el_{k-1} \vert}
\end{bmatrix}
S_k$}
\end{center}

\begin{ex}
\mbox{\scriptsize
$\els{X} = \texttt{abbc}
\begin{bmatrix}
\texttt{ab} \\ \texttt{aab} \\ \texttt{acca}
\end{bmatrix}
\texttt{cca}
\begin{bmatrix}
\texttt{aabcab} \\ \texttt{cba}
\end{bmatrix}
\texttt{bb}$} is an elastic-degenerate string, where we have the following:
\begin{itemize}
\item Three seeds: $S_1 =$ \mbox{ \scriptsize $\texttt{abbc}$, $S_2=\texttt{cca}$}, and $S_3=$ \mbox{\scriptsize $\texttt{bb}$}.

\item Two elastic-degenerate symbols: \\\mbox{\scriptsize $\el_1 = \begin{bmatrix}
\texttt{ab} \\ \texttt{aab} \\ \texttt{acca}
\end{bmatrix}$} and 
\mbox{\scriptsize $\el_2 = \begin{bmatrix}
\texttt{aabcab} \\ \texttt{cba}
\end{bmatrix}$}.

\item For $\el_1$: $E_{1,1} =$ \mbox{\scriptsize $\texttt{ab}$}, $E_{1,2} =$ \mbox{\scriptsize $\texttt{aab}$}, $E_{1,3} =$ \mbox{\scriptsize $ \texttt{acca}$}; minimum length is $2$ (length of $E_{1,1}$ is shortest); and  maximum length is $4$ (length of $E_{1,3}$ is longest).

\item For $\el_2$: $E_{2,1} =$ \mbox{\scriptsize $\texttt{aabcab}$}, $E_{2,2} =$ \mbox{\scriptsize $\texttt{cba}$}; minimum length is $3$ (length of $E_{2,1}$ is shortest); and maximum length is $6$ (length of $E_{2,1}$ is longest).
\end{itemize}
\end{ex}

Observe the use of $\els{X}$ to distinguish an elastic degenerate string from a (plain) solid string $X$ or a degenerate string $\tilde{X}$. In the following, we define three characteristics of a given elastic degenerate string $\els{X}$ (with $k$ seeds).

%we will be using $\el$ to denote an elastic-degenerate factor and $E_{i,j}$ to denote a string from the set representing elastic-degenerate factor $\el_i$ in a string. Further, 

\begin{definition}[Total Size: $\sz \els{X} \sz$]
\textit{Total size} of $\els{X}$, represented as $\sz \els{X} \sz$, is defined as the sum of the total length of its seeds and the total length of all the strings in each of its elastic-degenerate symbols (i.e. $\sz \els{X} \sz = \sig_{i=1}^{k} \vert S_i\vert + \sig_{i=1}^{k-1} \sig_{j=1}^{\vert \el_i \vert} \vert E_{i,j} \vert$).
\end{definition}

\begin{definition}[Length: $\vert \els{X} \vert$] The \textit{length} of $\els{X}$ is denoted by $\vert \els{X} \vert $ and is defined as the sum of the total length of its seeds and the total number of its elastic-degenerate symbols (i.e. $\vert \els{X} \vert = \sig_{i=1}^{k} \vert S_i\vert + k-1$. Informally, the total number of positions in $\els{X}$ is its length (considering an elastic-degenerate symbol to occupy only one position). Intuitively, a position belonging to some seed will be called \textit{solid position} and that of an elastic-degenerate symbol will be called \textit{elastic-degenerate position}. 
\end{definition}
In the running example, the total length of the seeds is $9$; hence, $\sz \els{X} \sz = 9 + (2+3+4) + (6+3) = 27$, while  $\vert \els{X} \vert = 9 + 2 = 11$. The first $\texttt{a}$ occurs at  (solid) position $1$, followed by $\texttt{b}$ at (solid) position $2$ and so on; $\el_1$and $\el_2$ are at (elastic-degenerate positions) $4$ and $9$, respectively; the last $\texttt{b}$ is at (solid) position $11$.

\begin{definition}[Possibility-Set: $\Re$] The \textit{possibility-set} $\Re$ of $\els{X}$ is a set of all possible (plain) \textit{solid} strings obtained from $\els{X}$. A solid string can be obtained by replacing each of the elastic-degenerate symbols with one of its constituent strings. More formally, it can be defined as follows:
\begin{center}
\mbox{$\Re = \{ S_1 E_{1,r_1} S_2 E_{2,r_2} \dd E_{k-1, r_{k-1} S_k}\} ~~\forall r_i, 1 \leq i \leq k-1$ such that $1 \leq r_i \leq \vert\el_i\vert $}
\end{center}
\end{definition}
For instance, in the running example,
$\Re =$ \mbox{\scriptsize $\{\texttt{abbc\underline{\texttt{ab}}cca\underline{aabcab}bb}, \texttt{abbc\underline{ab}cca\underline{cba}bb},$} \\ \mbox{\scriptsize 
$\texttt{abbc\underline{aab}cca\underline{aabcab}bb}, 
\texttt{abb\underline{aab}cca\underline{cba}bb},
\texttt{abbc\underline{acca}cca\underline{aabcab}bb}, 
\texttt{abbc\underline{acca}cca\underline{cba}bb}\}$} (constituent strings replacing the elastic-degenerate symbols have been underlined for clarity).

Now we can define \textit{matching} and \textit{occurrence} in the context of elastic-degenerate strings.

\begin{definition}[Matching] A given elastic-degenerate string $\els{X}$ is said to match a solid string $Y$ if, and only if, $Y \in \Re$ of $\els{X}$. It is represented as $\els{X} \match Y$. Informally, if one of the possible strings obtained from $\els{X}$ is the same as $Y$, we say that they match.

Analogously, two given elastics-degenerate strings $\els{X}$ and $\els{Y}$ match (represented as $\els{X} \match \els{Y}$) if and only if $\els{X} \cap \els{Y} \neq \phi$. Stating informally, both $\els{X}$ and $\els{Y}$ must produce at least one common solid string.
\end{definition}

\begin{ex}
Consider $\els{X}$ as given in  Example 1. If a string $Y =$ \mbox{\scriptsize$ \texttt{abbcabccacbabb}$} then $\els{X} \simeq Y$ whereas for a string $Z=$ \mbox{\scriptsize $\texttt{abbccccca}$}, $\els{X} \not \simeq Z$ as $Z$ does not occur in the possibility-set $\Re$ of $\els{X}$.
Given an elastic-degenerate string $\els{Y} =$ \mbox{\scriptsize $\texttt{ab} 
\begin{bmatrix}
\texttt{bcab} \\ \texttt{abb}
\end{bmatrix}
\texttt{ccacbabb}$}, $\els{X} \simeq \els{Y}$ as \mbox{\scriptsize$\texttt{abbcabccacbabb}$} is a common solid string obtained from both.
\end{ex}

\begin{definition}[Occurrence] 
Given two positions $i$ and $j$ in an  elastic-degenerate string (text) $\els{T}$, let $S$ be some solid string obtained from $\els{T}[i\dd j]$ (i.e. $S\in \Re$ ). A given solid string (pattern) $P$ is said to occur in $\els{T}$ between positions $i$ and $j$, if 
\begin{center}
$
\begin{cases}
P = S & \text{if both, $i$ and $j$, are solid} \\
P \text{ is prefix of } S & \text{if $i$ is solid and $j$ is elastic-degenerate} \\
P \text{ is suffix of } S & \text{if $i$ is elastic-degenerate and $j$ is solid} \\
P \text{ is factor of } S & \text{if both, $i$ and $j$, are elastic-degenerate} \\
\end{cases}
$
\end{center}
An occurrence is represented as the pair of start-position (head) and end-position (tail).
\end{definition}
For consistency with the intuitive meaning of an occurrence, we say that $P$ occurs at the position of some elastic-degenerate symbol (say $\el_i$) of $\els{T}$, if it is a factor of any of the constituent strings of $\el_i$.

\begin{ex}
Consider a pattern $P =$ \mbox{\scriptsize $\texttt{cabbcb}$} and a text $\els{T}$  as follows:
\begin{center}
\mbox{\scriptsize
$
\texttt{aacabbcbbc}
\begin{bmatrix}
\texttt{a} \\ \texttt{aab} \\ \texttt{acca}
\end{bmatrix}
\texttt{bb}
\begin{bmatrix}
\texttt{c} \\ \texttt{acabbcbb} \\ \texttt{cba}
\end{bmatrix}
\texttt{bacabbc}
\begin{bmatrix}
\texttt{b} \\ \texttt{cabb} \\ \texttt{bbc} \\ \texttt{aacabb}
\end{bmatrix}
\texttt{cbc}
$ }
\end{center}
All the occurrences of $P$ in $\els{T}$ are given in Table~\ref{table:Ex3}.

\begin{table}[h]
\centering
\scriptsize
\begin{tabular}{||l||c|c|c|c|c|c|c||}
\hline\hline
 & & & & &  &  &  \\
 \textbf{Occurrence:} & $(3,8)$ & (10,15) & (11,14) & (11,15) & (14,14) & (17,22) & (22,24) \\ \hline
 & & & & &  &  &  \\
 &  & $\el_1$: $\underline{\texttt{a}}$ & $\el_1$: $\texttt{ac}\underline{\texttt{ca}}$ & $\el_1$: $\texttt{ac}\underline{\texttt{ca}}$ & $\el_2$: $\texttt{ac}\underline{\texttt{cabbbc}}\texttt{b}$ & $\el_3$: $\underline{\texttt{c}}$ & $\el_3$: $\underline{\texttt{cabb}}$ \\

\textbf{Strings chosen:}& - & & & &  &  &  \\

                 &  &  $\el_2$: $\underline{\texttt{c}}$ & $\el_2$: $\underline{\texttt{cb}}\texttt{a}$ & $\el_2$: $\underline{\texttt{c}}$ & &or $\el_3$: $\underline{\texttt{c}}\texttt{bc}$ & or $\el_3$: $\texttt{aa}\underline{\texttt{c}}$  \\ 
& & & & &  &  &  \\\hline \hline
\end{tabular}
\caption{Table representing the occurrences of $P$ in $\els{T}$ as given in Example $3$.}
\label{table:Ex3}
\end{table}
\end{ex}
\begin{comment}

\end{comment}
Note that more than one occurrence of $P$ can start from the same position but their ending-positions are different (for instance, $(11,14)$ and $(11,15)$ in Example $3$). Also, note that different strings in the same elastic-degenerate symbols can lead to the same occurrence i.e. same pair of head and tail (as happened for occurrences $(17,22)$  and $(17,24)$ in Example $3$).

\begin{ex}
Here, we illustrate the case, where an elastic-degenerate string has an empty string as a seed. Consider 
\begin{center}

$\els{T}  =$ \mbox{\scriptsize $\texttt{ab} 
\begin{bmatrix}
\texttt{bcab} \\ \texttt{abb}
\end{bmatrix}
\begin{bmatrix}
\texttt{ab} \\ \texttt{cbb} \\ \texttt{abc}
\end{bmatrix}
\texttt{cca}
\begin{bmatrix}
\texttt{bb} \\ \texttt{cb}
\end{bmatrix}
\texttt{ca}$} and 
a pattern $P =$ \mbox{\scriptsize $\texttt{babbcb}$},
\end{center}
there is an occurrence of $P$ at $(2, 4)$ of $\els{T}$. 
\end{ex}

%---------%---------%---------%---------%---------%---------%---------%--------%
%---------%---------%---------%---------%---------%---------%---------%--------%
%---------%---------%---------%---------%---------%---------%---------%--------%
\section{Algorithmic Tools}
\label{sec:tools}

Here, we briefly introduce a fundamental data structure, which supports a wide variety of string matching algorithms, and a well-known pattern matching algorithm. This data structure and pattern matching algorithm will be used by the proposed algorithm.

%---------%---------%---------%---------%---------%---------%---------%--------%
%---------%---------%---------%---------%---------%---------%---------%--------%
\subsection*{Suffix Tree} The \textit{suffix tree} $\mathcal{S}(X)$ of a non-empty string $X$ of length $n$, is a compact trie representing all the suffixes of $X$ such that $\mathcal{S}(X)$  has $n$ leaves, labelled from $1$ to $n$.
%Additionally, each edge is labelled with a letter of $\Sigma$. For any $i, 1 \leq i \leq n$, the concatenation of the edge labels on the path from the root of $\mathcal{S}(X)$  to leaf $i$ is precisely the suffix $X[i\dd n]$. For any two suffixes $U = X[i\dd n]$ and  $V= X[j\dd n]$ of $X$, if $W$ is the longest common prefix of $U$ and $V$, then the path in $\mathcal{S}(X)$  corresponding to $W$ is the same for $U$ and $V$. In other words, length of the Least Common Ancestor (LCA) of the two leaves is same as the LCP of the suffixes represented by those leaves.
For a general introduction to suffix trees, see~\cite{CHL07cup}. 
%If $v$ is a node of $\mathcal{S}(X)$, then the \textit{path-label} of $v$ is  the concatenation of  the edge labels along the path from the root to $v$. We define the \textit{leaf-list} of $v$ as a list of the leaf labels in the subtree below $v$.
The construction of the suffix tree $\mathcal{S}(X)$ of the input string $X$  takes  $\mathcal{O}(n)$ time and space, for string over a fixed-sized alphabet \cite{weiner1973linear,mccreight1976space,ukkonen1995line}. Once the suffix tree of a given string (called text) has been constructed, it can be used to support queries that return the occurrences of a given string (called pattern) in time linear in the length of the pattern.  Least Common Ancestor (LCA) of the two leaves of a suffix tree can be computed in constant time after a linear time preprocessing to answer LCA queries \cite{Harel1984,Schieber:1988:FLC:55258.55270}.
A generalised suffix tree is a suffix tree for a set of strings
%It can be obtained, for a given set of $l$ strings $X = \{X_1, X_2, \cdots, X_l\}$, by constructing the suffix tree of the concatenated string $X_1\$_1 X_2\$_2 \cdots X_k\$_l$, where each $\$_i ~ \forall i \in [1 \cdots k]$ is unique end-marker for each string $X_i ~ \forall i \in [1 \cdots k]$
%distinct and does not occur anywhere in any of the given strings. 
\cite{AMIR1994208,Gusfield:1997:AST:262228}.

\subsection*{KMP Algorithm and failure function}
Knuth, Morris and Pratt (KMP) discovered the first linear time string-matching algorithm \cite{doi:10.1137/0206024}, that is the problem of finding all occurrences of a pattern $P$ in a text $T$. The KMP algorithm follows the na\"{\i}ve approach for this problem, that is,  it slides the pattern across the text. Additionally it preprocesses the pattern $P$ by computing a \textit{failure function} $f$ that indicates the largest possible shift, using previously performed comparisons. Specifically, the \textit{failure function} 
$f(i)$ is defined as the length of the longest prefix of $P$ that is a suffix of $P[1..i]$. By using the failure function, it achieves an optimal search time of $O(n)$ after $O(m)$-time pre-processing, where $n$ is the length of $T$ and $m < n$ is the length of $P$.
%\newpage

%---------%---------%---------%---------%---------%---------%---------%--------%
%---------%---------%---------%---------%---------%---------%---------%--------%
%---------%---------%---------%---------%---------%---------%---------%--------%
%---------%---------%---------%---------%---------%---------%---------%--------%
%%%%%%%%%%%%%%%% Section 4 %%%%%%%%%%%%%%%%%%%%%%%%%%%%%5

\section{Algorithm for pattern matching in elastic-degenerate text}
\label{sec:algo}

%Here, we extend the classical pattern matching problem in the context of elastic-degenerate strings.

\subsection{Problem Definition}
\label{caseone}

\defproblem{\textsc{Problem: Finding Occurrences in Elastic-Degenerate Text given a Solid  Pattern}}
{A pattern $P$ of length $m$, an elastic-degenerate text $\els{T} = S_1 \el_1 S_2 \dd \el_{k-1} S_k$, of length $n$ and total size $N$, where each $\el_i = \{E_{i,j}\}, ~ 1 \leq j \leq \vert\el_i\vert$.}
{All the occurrences of $P$ in $\els{T}$.}

All the occurrences of the pattern $P$ in the text $\els{T}$, fall under the following cases:
%There are only the following possibilities, for an occurrence of the pattern in the text:
\begin{enumerate}
\item $P$ entirely lies in some seed $S_i$.
\item $P$ entirely lies in some string of an elastic-degenerate symbol $\el_i$.
\item $P$ spans across one or more elastic-degenerate symbols. This can further be seen as:
\begin{enumerate}
\item $P$ starts in some seed $S_i$.
\item $P$ begins in some string of an elastic-degenerate symbol $\el_i$.
\end{enumerate}
\end{enumerate}

For instance, consider Example $3$ - the occurrences $(3,8)$ and $(14,14)$ lie in Case $1$ and Case $2$, respectively; $(10,15)$ and $(17,22)$ belong to Case $3(a)$; Case $3(b)$ covers $(11,14)$, $(11,15)$, and $(22,24)$.

\subsection{Algorithm}
We now present an efficient algorithm that makes use of the KMP pattern matching algorithm and the suffix tree. Clearly, KMP pattern matching algorithm can easily report the occurrences corresponding to the Cases $1$ and $2$. Case $3$ requires some additional processing and data-structures. The algorithm works in two stages, outlined in the following:

\subsubsection*{Stage 1: Pre-processing}
Preprocess the pattern $P$ to compute its failure-function as required for KMP algorithm. In addition, create a generalised suffix tree $\mathcal{ST}_{S_i}$ for the set $\{P, S_i\}$ corresponding to each seed $S_i$, $1 \leq i \leq k$, as well as a generalised suffix tree $\mathcal{ST}_{\el_i}$ for the set $\{P, E_{i,1}, E_{i,2},\dd, E_{i,\vert \el_i\vert}\}$ corresponding to each elastic-degenerate symbol $\el_i$, $1 \leq i \leq k-1$. Furthermore, pre-process these suffix trees so as to answer the longest common ancestor (LCA) queries in constant time.

\subsubsection*{Stage 2: Search}
Start searching the pattern in the text using the KMP algorithm, comparing the letters and using failure function to shift the pattern on a mismatch. 
The starting position of an occurrence being tested may be either solid or elastic-degenerate; we call the two types of occurrences as  \textit{Type $1$} and \textit{Type $2$}, respectively. We consider the two types separately as follows:

\mystep{Type $1$: Solid start-position}
Consider a situation, where an occurrence starting from a position (say $pos$) that lies in some seed $S_i$, is being tested. Proceed normally comparing the corresponding letters of $P$ and $S_i$; shifting the pattern using failure function on mismatch. As soon as the elastic-degenerate symbol $\el_i$ is encountered (suppose corresponding position in the pattern is $p$), abort the KMP algorithm (for this test). Check each of the strings of $\el_i$ (i.e. $E_{i,j}$) whether or not it occurs in the pattern at position $p$, using LCA queries on $\mathcal{ST}(\el_i)$; \textit{ticking} (marking) the tails of the found occurrences. It can be realized by maintaining a boolean array of size $m$, called $\ticked_i$. 

Next, Procedure~\ref{A1} (given below) is followed; in which each ticked position of $\ticked_i$ is tried to extend by testing whether $S_{i+1}$ occurs adjacent to it (using LCA queries on $\mathcal{ST}_{S_{i+1}}$). For each such found occurrence of $S_{i+1}$, occurrences of strings of $\el_{i+1}$ are checked using the suffix tree $\mathcal{ST}_{\el_{i+1}}$ and their tails are ticked in $\ticked_{i+1}$. The procedure will then be repeated for $\ticked_{i+1}$; it continues recursively until there is no tail marked in some call. 

Once the process ends (reporting all the occurrences of $P$ starting from $pos$, if any), the failure function corresponding to the position where the KMP algorithm was aborted (i.e. $p$) is used to shift the pattern and the KMP algorithm resumes. 

\begin{figure}[!h]
\begin{tikzpicture}[xscale=0.8, yscale=0.6]
\path [fill=mygray] (0,3.3) rectangle (0.97,2.7);
\draw (0,3.2) -- (0,2.8);
\node[align=center, above] at (0,3.4)%
{$pos$};

\draw [thick, dashed] (-2,3) -- (-1,3);
\node[align=center, left] at (-2,3)%
{$\els{T}$};
%el
\foreach \x/\y in 	{{1/2.5}, {4.5/6}, {7/8.5}}
{
\draw [thick] (\x,4.5) -- (\x,1.5);
\draw [thick] (\y,4.5) -- (\y,1.5);
}
%seed
\foreach \x/\y in 	{{-1/1}, {2.5/4.5}, {6/7}, {8.5/9.5}}
{
\draw [thick] (\x,3) -- (\y,3);
}
\draw [thick, dashed] (9.5,3) -- (12,3);

%label
\foreach \name/\x in 	{{$S_i$/-0}, {$\el_i$/1.75}, {$S_{i+1}$/3.5}, {$\el_{i+1}$/5.25},{$S_{i+2}$/6.5}, {$\el_{i+2}$/7.75},{$S_{i+3}$/9}}
{
\node[align=center, above] at (\x,4.6)%
{\name};
}
%eij
\foreach \name/\x in {{$E_{i,\vert\el_{i}\vert}$/1.75}, {$E_{i+1,\vert\el_{i+1}\vert}$/5.25}, {$E_{i+2,\vert\el_{i+2}\vert}$/7.75}}
{
\node[align=center, above] at (\x,1.5)%
{\tiny{\name}};
}
\foreach \name/\x in {{$E_{i,1}$/1.75}, {$E_{i+1,1}$/5.25}, {$E_{i+2,1}$/7.75}}
{
\node[align=center, below] at (\x,4.5)%
{\tiny{\name}};
}
\foreach \name/\x/\y in {{$E_{i,r_1}$/1.75/3.4}, {$E_{i,r_2}$/1.75/2.5},{$E_{i+1,j_1}$/5.25/3.6}, {$E_{i+1,j_2}$/5.25/3}, {$E_{i+1,j_3}$/5.25/2.4},{$E_{i+2,p_1}$/7.75/3.4},{$E_{i+2,p_2}$/7.75/2.6}}
{
\node[align=center] at (\x,\y)%
{\tiny{\name}};
}
\draw [dotted] (1.75,2) -- (1.75,4.1);
\draw [dotted] (5.25,2) -- (5.25,4.1);
\draw [dotted] (7.75,2) -- (7.75,4.1);

%%%%%Initial
\path [fill=mygray] (0,0.3) rectangle (1,-0.3);
\draw [thick] (0,0) -- (10.7,0);
\node[align=center, left] at (0,0)%
{$P$};

%S1: 1 : -0.5
\draw [thick,decorate,decoration={brace}]
(-1, -1.6) -- (-1,-0.6) node [black,midway,left, xshift=-10pt] {$\ticked_i$};

\draw (0,-0.7) -- (1, -0.7);
\draw (1,-.2) -- (1, .2);
\node[align=center, above] at (1,.2)%
{$p$};

%Ti
\draw[zig](1,-0.7) -- (2, -0.7);
\node[align=center, above] at (1.5,-0.7)%
{$E_{i,r_1}$};
\node[dot] at (2,0){};%
\node[align=center, above] at (2,.2)%
{$e_1$};
\node[align=center, right] at (2,-0.7)%
{$\checkmark$};

\draw[zig](1,-1.5) -- (3, -1.5);
\node[align=center, above] at (2,-1.5)%
{$E_{i,r_2}$};
\node[dot] at (3,0){};%
\node[align=center, above] at (3,.2)%
{$e^\prime_1$};
\node[align=center, right] at (3,-1.5)%
{$\checkmark$};

%Ti+1
\draw [thick,decorate,decoration={brace}]
(-1, -4.9) -- (-1,-2.9) node [black,midway,left, xshift=-10pt] {$\ticked_{i+1}$};

\draw(2,-3) -- (4, -3);
\node[align=center, above] at (3,-3)%
{$S_{i+1}$};
\draw[zig](4,-3) -- (5.5, -3);
\node[align=center, above] at (4.75,-3)%
{$E_{i+1,j_1}$};
\node[dot] at (5.5,0){};%
\node[align=center, above] at (5.5,.2)%
{$e_2$};
\node[align=center, right] at (5.5,-3)%
{$\checkmark$};

\draw(3,-4) -- (5, -4);
\node[align=center, above] at (4,-4)%
{$S_{i+1}$};

\draw[zig](5,-4) -- (6, -4);
\node[align=center, above] at (5.5,-4)%
{$E_{i+1,j_2}$};
\node[dot] at (6,0){};%
\node[align=center, above] at (6,.2)%
{$e^\prime_2$};
\node[align=center, right] at (6,-4)%
{$\checkmark$};

\draw[zig](5,-4.8) -- (7.5, -4.8);
\node[align=center, above] at (6.25,-4.8)%
{$E_{i+1,j_3}$};
\node[dot] at (7.5,0){};%
\node[align=center, above] at (7.5,.2)%
{$e^{\prime\prime}_2$};
\node[align=center, right] at (7.5,-4.8)%
{$\checkmark$};

%Sti+2
\draw [thick,decorate,decoration={brace}]
(-1, -8.1) -- (-1,-5.9) node [black,midway,left, xshift=-10pt] {$\ticked_{i+2}$};

\draw(5.5,-6) -- (6.5, -6);
\node[align=center, above] at (6,-6)%
{$S_{i+2}$};
\node[align=center, right] at (6.5,-6)%
{\textbf{X}};

\draw(6,-7) -- (7, -7);
\node[align=center, above] at (6.5,-7)%
{$S_{i+2}$};
\draw[zig](7,-7) -- (9.7, -7);
\node[align=center, above] at (8.35,-7)%
{$E_{i+2,p_1}$};
\node[dot] at (9.7,0){};%
\node[align=center, above] at (9.7,.2)%
{$e^\prime_3$};
\node[align=center, right] at (9.7,-7)%
{$\checkmark$};

\draw(7.5,-8) -- (8.5, -8);
\node[align=center, above] at (7.85,-8)%
{$S_{i+2}$};
\draw[zig](8.5,-8) -- (9.3, -8);
\node[align=center, above] at (8.95,-8)%
{$E_{i+2,p_2}$};
\node[dot] at (9.3,0){};%
\node[align=center, above] at (9.3,.2)%
{$e_3$};
\node[align=center, right] at (9.3,-8)%
{\textbf{X}};

%Ti+3
\draw [thick,decorate,decoration={brace}]
(-1, -9.3) -- (-1,-8.7) node [black,midway,left, xshift=-10pt] {$\ticked_{i+3}$};

\draw(9.7,-9) -- (10.7, -9);
\node[align=center, above] at (10.1,-9)%
{Prefix($S_{i+4}$)};
\draw (10.7,.5) -> (10.7, .2);
\node[align=center, above] at (10.7,.7)%
{pattern ends};

\end{tikzpicture}
\caption{An illustration of how the algorithm worked as described in Type $1$. Strings in elastic-degenerate symbols have been shown as zigzag, while solid lines depict the seeds. Symbol \textbf{X} denotes that this path could not be extended further while the symbol $\checkmark$ represents a ticked tail.} 
\label{fig:sit_1}
\end{figure}
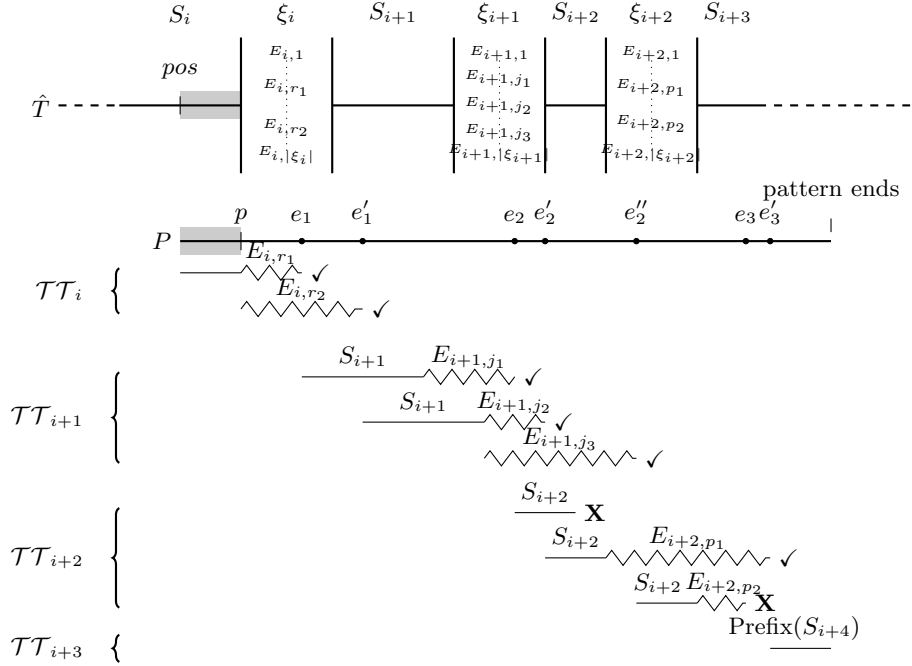

It is to be noted that an occurrence of $P$ is implied, if the length of LCA of the pattern starting from some ticked-tail $t$ with either of the following hits the boundary of the pattern:
\begin{itemize}
\item some seed $S_j$ (i.e. $\vert LCA_{t, S_j} \vert + t > m$)
\item any string $E_{i,j}$ of some elastic-degenerate symbol $\el_i$ (i.e. $\vert LCA_{t, E_{i,j}} \vert + t > m$).
\end{itemize}
Figure~\ref{fig:sit_1} elucidates the description given above.

\begin{algorithm2e}%[H]
\scriptsize
\SetAlgorithmName{Procedure}{}
\DontPrintSemicolon
\SetKwData{Flag}{$isNonEmpty$}
\SetKwData{LCA}{l}
\SetKwData{LCA}{l}
\SetKwData{OcS}{$\mymat{Occ}(S)$}
\SetKwData{Tick}{t}
\SetKwData{Tail}{e}
\SetKwFunction{Extend}{Extend}
\SetKwFunction{LCAQuery}{LCA}

\SetKwInOut{Input}{input}\SetKwInOut{Output}{output}
\Input{A boolean array $\ticked_i$ of size $m$ indicating ticked tails to be extended.}
\Output{Reporting the found occurrences and preparing $\ticked_{i+1}$ for the next recursive call.}
%\Act $\leftarrow 1$\; %\tcp*[r]{Last position of current clump}
\BlankLine
\Flag $\leftarrow ~false$;
\BlankLine
\ForAll{$\Tick$ \KwIn $\ticked_i$ which are ticked}{
	$\LCA_s \leftarrow \vert$  \LCAQuery{$P[\Tick+1 \dd m]$, $S_{i+1}[1 \dd \vert S_{i+1} \vert]$} $\vert$;
    
    \If(\tcp*[f]{Pattern ends}){$(\LCA_s + \Tick) > m$}{
        Report the occurrence;
	}
	\ElseIf(\tcp*[f]{$S_{i+1}$ occurs here}){$\LCA_s = \vert S_{i+1} \vert$}{
    	$\Tail \leftarrow \Tick+\vert S_{i+1} \vert$;
        
    	\ForAll{$E_{i+1,j}$ in $\el_{i+1}$}{
        	$\LCA_e \leftarrow \vert$  \LCAQuery{$P[\Tail \dd m]$, $E_{i+1,j}[1 \dd \vert E_{i+1,j} \vert]$} $\vert$;
            
            \If(\tcp*[f]{Pattern ends}){$(\LCA_e + \Tail) > m$}{
        		Report the occurrence (if not reported already);
			}
        	\ElseIf(\tcp*[f]{$E_{i+1,j}$ occurs here}){$\LCA_e = \vert E_{i+1,j} \vert$}{
        		Mark $\Tail + \vert E_{i+1,j} \vert - 1$ in $\ticked_{i+1}$;
                
                \Flag $\leftarrow ~true$;
			}
        }
        
	}
}
\BlankLine
\If{\Flag}{
\Extend{$\ticked_{i+1}$};
}

%}
\caption{Procedure to extend ticked tails in a given $\ticked_{i}$ and reporting the occurrences found, if any.\label{A1}}
\end{algorithm2e}

%Clearly, the method described above reports all the occurrences corresponding to Case $3a$. 
\mystep{Type $2$: Elastic-Degenerate start-position}
Now consider a situation, where the starting position of an occurrence to be tested is an elastic-degenerate symbol $\el_i$. This case can be processed in the similar fashion as one described for Type $1$, with the only difference in the manner in which tails are ticked initially. 

Begin by applying the KMP algorithm for each $E_{i,j}$, to achieve two purposes: finding the occurrences of $P$ in $E_{i,j}$ and ticking the last position of $E_{i,j}$ for which a prefix of $P$ appears as a suffix of $E_{i,j}$. The ticked tails obtained in that way, are then extended by Procedure~\ref{A1} recursively and occurrences are reported, if any. After the Procedure~\ref{A1} ends, the KMP algorithm resumes and testing of the pattern starts at the beginning of the seed $S_{i+1}$.  

%\subsection{Practical details}

%---------%---------%---------%---------%---------%---------%---------%--------%
%---------%---------%---------%---------%---------%---------%---------%--------%

\section{Analysis}
\label{sec:analysis}
In this section, we discuss the correctness of the algorithm and analyse its space and time complexity.
\subsection{Correctness}
Correctness of the presented algorithm is straightforward as every position of the text is being tested for an occurrence exhaustively. While the occurrences corresponding to the Cases $1$ and $3(a)$ are covered by Type $1$, Type $2$ investigates all the occurrences associated with Case $2$ and Case $3(b)$ - thus all the occurrences of $P$ in $\els{T}$ are reported.

\subsection{Space Complexity}
The space needed by both, the failure-function and ticked tails array, is $\mathcal{O}(m)$. Each suffix tree $\mathcal{ST}_{S_i}$ uses $\mathcal{O}(m + \vert S_i \vert)$ and  $\mathcal{ST}_{\el_i}$ takes $\mathcal{O}(m + \sig_{i=1}^{k-1} \sig_{j=1}^{\vert \el_i \vert} \vert E_{i,j} \vert)$ space, leading to the total space occupied by the tree to be $\mathcal{O}(km + N)$. Thus, assuming $k$ to be constant, the solution only needs the space that is linear in the input size.

An important fact that can be exploited to make the algorithm further space-efficient is that all the suffix trees are not required in the memory at the same time. Once the start-position crosses past a seed or an elastic-degenerate symbol, their corresponding trees are no longer needed and can be discarded.

\subsection{Time complexity}
Time taken by the preprocessing stage is $\mathcal{O}(km + N)$ as the failure function can be computed in $\mathcal{O}(m)$ time and construction of all the suffix trees (along with their preprocessing required to answer LCA queries in constant time) can be done in  $\mathcal{O}(km + N)$ time.

The search stage uses the KMP algorithm over each seed and each string of every elastic-degenerate symbol in the text, to report the occurrences for Case $1$ and Case $2$ and to search the beginning of the occurrence for Case $3$. Thus, overall the time consumed by the KMP algorithm is $\mathcal{O}(\sig_{i=1}^{k} \vert S_i\vert + \sig_{i=1}^{k-1} \sig_{j=1}^{\vert \el_i \vert} \vert E_{i,j} \vert)$  (i.e. $\mathcal{O}(N)$).

Procedure~\ref{A1} can be analysed as follows: Intuitively, for every ticked position in the pattern (which can at most be $m$), LCP is calculated (in constant time) to find whether the corresponding seed occurs at the ticked position or not; a found such occurrence  is then tried to extend by computing LCP with each of the strings in the following elastic-degenerate symbol. If $\alpha$ is the largest number of strings in any elastic-degenerate symbol of the text, this extension-step for each ticked position will  be carried out at most $\alpha$ times. More specifically, the outer loop of the procedure runs $m$ times and the inner one takes $\ord(\alpha)$ time, as each LCA query takes constant time. Thus, each recursive call requires $\ord(m\alpha)$ time. The number of recursive calls depends on the number of the elastic-degenerate symbols spanned by the occurrence of $P$ being tested. In other words, if an occurrence spans across $i$ elastic-degenerate symbols, there will be $i$ recursive calls to the procedure. If $\gamma$ is the maximum such $i$, Procedure~\ref{A1} executes in $\ord(m\alpha \gamma)$ time (in total) for each start-position.
%Procedure~\ref{A1} can be analysed as follows: The outer loop of the procedure runs $m$ times and the inner one takes $\ord(\alpha)$ time where $\alpha$ is the maximum number of strings in some elastic-degenerate symbol, as each LCA query takes constant time. Thus, each recursive call requires $\ord(m\alpha)$ time. The number of recursive calls depends on the number of the elastic-degenerate symbols spanned by the occurrence of $P$ being tested. In other words, if an occurrence spans across $i$ elastic-degenerate symbols, there will be $i$ recursive calls to the procedure. If $\gamma$ is the maximum such $i$, Procedure~\ref{A1} executes in $\ord(m\alpha \gamma)$ time (in total) for each start-position. 

Initial ticking of the tails in Type $1$ needs $\mathcal{O}(\alpha)$ time. For Type $2$, initial ticking is done by KMP algorithm (already accounted above). In the worst case, Procedure~\ref{A1} will be called from each of the $n$ positions of the text, leading to an overall time-complexity of the algorithm to be $\ord(nm\alpha \gamma + N)$ (as $k \leq n$). In real data, $\alpha$ and $\gamma$ are mostly very small constants. Therefore, the algorithm is expected to work really efficiently in practice.

%---------%---------%---------%---------%---------%---------%---------%--------%
%---------%---------%---------%---------%---------%---------%---------%--------%

\section{Conclusion}
\label{sec:conclusion}
Motivated by the applications in genomics, we extended the notion of gapped strings to elastic-degenerate strings in this paper. We presented an efficient algorithm for the pattern matching problem, given a (solid) pattern and an elastic-degenerate text, running in $\ord(N+\alpha\gamma nm)$ time; where $m$ is the length of the given pattern; $n$ and $N$ are the length and total size of the given elastic-degenerate text, respectively; $\alpha$ and $\gamma$ are small constants, respectively representing the maximum number of strings in any elastic-degenerate symbol of the text and the largest number of elastic-degenerate symbols spanned by any occurrence of the pattern in the text. Note that $\alpha$ and $\gamma$ are so small in real-world applications that the algorithm is expected to work very efficiently in practice. The space used by the algorithm is linear in the size of the input for a constant number of elastic-degenerate symbols in the text.

It is to be noted that the presented algorithm can easily be adapted for a conserved (simple) degenerate pattern by using the algorithm given in \cite{Crochemore2016109} for conserved degenerate pattern matching. An interesting further direction is to conduct large-scale experiments, specifically in the context of studying inter-species genetic-variations~\footnote{A proof-of-concept implementation of our algorithm can be accessed at \url{https://github.com/Ritu-Kundu/ElDeS}. Due to lack of space, experimental results are not included in the current version; they will be added in the full version of the paper.}. Furthermore, other domains that involve web-mining applications may find the presented solution interesting and beneficial.
\bibliographystyle{splncs03}
\bibliography{references}

\begin{thebibliography}{10}
\providecommand{\url}[1]{\texttt{#1}}
\providecommand{\urlprefix}{URL }

\bibitem{AMIR1994208}
Amir, A., Farach, M., Galil, Z., Giancarlo, R., Park, K.: Dynamic dictionary
  matching. Journal of Computer and System Sciences  49(2),  208 -- 222 (1994),
  \url{http://www.sciencedirect.com/science/article/pii/S0022000005800479}

\bibitem{Church2015}
Church, D.M., Schneider, V.A., Steinberg, K.M., Schatz, M.C., Quinlan, A.R.,
  Chin, C.S., Kitts, P.A., Aken, B., Marth, G.T., Hoffman, M.M., Herrero, J.,
  Mendoza, M.L.Z., Durbin, R., Flicek, P.: Extending reference assembly models.
  Genome Biology  16(1), ~13 (2015),
  \url{http://dx.doi.org/10.1186/s13059-015-0587-3}

\bibitem{CHL07cup}
Crochemore, M., Hancart, C., Lecroq, T.: Algorithms on Strings. Cambridge
  University Press (2007), 392 pages

\bibitem{Crochemore2016109}
Crochemore, M., Iliopoulos, C.S., Kundu, R., Mohamed, M., Vayani, F.: Linear
  algorithm for conservative degenerate pattern matching. Engineering
  Applications of Artificial Intelligence  51,  109 -- 114 (2016),
  \url{http://www.sciencedirect.com/science/article/pii/S0952197616000130},
  mining the Humanities: Technologies and Applications

\bibitem{5ab9058d631145c29eb4abcd7346305a}
Crochemore, M., Sagot, M.F.: Motifs in Sequences: Localization and Extraction,
  pp. 47--97. Marcel Dekker, New York (2004)

\bibitem{Dilthey2015}
Dilthey, A., Cox, C., Iqbal, Z., Nelson, M.R., McVean, G.: Improved genome
  inference in the mhc using a population reference graph. Nat Genet  47(6),
  682--688 (Jun 2015), \url{http://dx.doi.org/10.1038/ng.3257}, technical
  Report

\bibitem{Gusfield:1997:AST:262228}
Gusfield, D.: Algorithms on Strings, Trees, and Sequences: Computer Science and
  Computational Biology. Cambridge University Press, New York, NY, USA (1997)

\bibitem{Harel1984}
Harel, H.T., Tarjan, R.E.: Fast algorithms for finding nearest common
  ancestors. {SIAM} J. Comput.  13(2),  338--355 (1984)

\bibitem{Huang01072013}
Huang, L., Popic, V., Batzoglou, S.: Short read alignment with populations of
  genomes. Bioinformatics  29(13),  i361--i370 (2013),
  \url{http://bioinformatics.oxfordjournals.org/content/29/13/i361.abstract}

\bibitem{doi:10.1137/0206024}
Knuth, D.E., James H.~Morris, J., Pratt, V.R.: Fast pattern matching in
  strings. SIAM Journal on Computing  6(2),  323--350 (1977),
  \url{http://dx.doi.org/10.1137/0206024}

\bibitem{Liu2014}
Liu, Y., Koyut{\"u}rk, M., Maxwell, S., Xiang, M., Veigl, M., Cooper, R.S.,
  Tayo, B.O., Li, L., LaFramboise, T., Wang, Z., Zhu, X., Chance, M.R.:
  Discovery of common sequences absent in the human reference genome using
  pooled samples from next generation sequencing. BMC Genomics  15(1),  685
  (2014), \url{http://dx.doi.org/10.1186/1471-2164-15-685}

\bibitem{Maciuca2016}
Maciuca, S., del Ojo~Elias, C., McVean, G., Iqbal, Z.: A Natural Encoding of
  Genetic Variation in a Burrows-Wheeler Transform to Enable Mapping and Genome
  Inference, pp. 222--233. Springer International Publishing, Cham (2016),
  \url{http://dx.doi.org/10.1007/978-3-319-43681-4_18}

\bibitem{mccreight1976space}
McCreight, E.M.: A space-economical suffix tree construction algorithm. Journal
  of the ACM (JACM)  23(2),  262--272 (1976)

\bibitem{Pissis2014}
Pissis, S.P.: Motex-ii: structured motif extraction from large-scale datasets.
  BMC Bioinformatics  15(1),  235 (2014),
  \url{http://dx.doi.org/10.1186/1471-2105-15-235}

\bibitem{Rahman2006}
Rahman, M.S., Iliopoulos, C.S., Lee, I., Mohamed, M., Smyth, W.F.: Computing
  and Combinatorics: 12th Annual International Conference, COCOON 2006, Taipei,
  Taiwan, August 15-18, 2006. Proceedings, chap. Finding Patterns with Variable
  Length Gaps or Don't Cares, pp. 146--155. Springer Berlin Heidelberg, Berlin,
  Heidelberg (2006), \url{http://dx.doi.org/10.1007/11809678_17}

\bibitem{Schieber:1988:FLC:55258.55270}
Schieber, B., Vishkin, U.: On finding lowest common ancestors: Simplification
  and parallelization. SIAM J. Comput.  17(6),  1253--1262 (Dec 1988),
  \url{http://dx.doi.org/10.1137/0217079}

\bibitem{ukkonen1995line}
Ukkonen, E.: On-line construction of suffix trees. Algorithmica  14(3),
  249--260 (1995)

\bibitem{weiner1973linear}
Weiner, P.: Linear pattern matching algorithms. In: Proceedings of the 14th
  IEEE Annual Symposium on Switching and Automata Theory. pp. 1--11. Institute
  of Electrical Electronics Engineer (1973)

\end{thebibliography}

%---------%---------%---------%---------%---------%---------%---------%--------%

\end{document}